\documentclass{nature}

\usepackage{amssymb,amsmath,graphicx,longtable}
\def\simlt{\mathrel{\hbox{\rlap{\hbox{\lower4pt\hbox{$\sim$}}}\hbox{$<$}}}}
\def\simgt{\mathrel{\hbox{\rlap{\hbox{\lower4pt\hbox{$\sim$}}}\hbox{$>$}}}}

\def\ale{\mathrel{\hbox{\rlap{\hbox{\lower4pt\hbox{$\sim$}}}\hbox{$<$}}}}
\def\age{\mathrel{\hbox{\rlap{\hbox{\lower4pt\hbox{$\sim$}}}\hbox{$>$}}}}

\begin{document}

\title{Solar flare-like origin of X-ray
flares in gamma-ray burst afterglows}

\author{F. Y. Wang$^{1,2}$ \& Z. G. Dai$^{1,2}$}

\date{\today}{}
\maketitle

\begin{affiliations}
\item School of Astronomy and Space Science, Nanjing
University, Nanjing 210093, China   %
\item Key Laboratory of
Modern Astronomy and Astrophysics (Nanjing University),
Ministry of Education, Nanjing 210093, China %
\end{affiliations}

\begin{abstract}
X-ray flares detected in nearly half of gamma-ray burst (GRB)
afterglows are one of the most intriguing phenomena in high-energy
astrophysics\cite{f1,f2,f3,f4,f5,f6,f7,f8}. All the observations
indicate that the central engines of bursts, after the gamma-ray
emission has ended, still have long periods of activity, during
which energetic explosions eject relativistic materials, leading to
late-time X-ray emission\cite{f2,f9,z07}. It is thus expected that
X-ray flares provide important clues to the nature of the central
engines of GRBs, and more importantly, unveil the physical origin of
the flares themselves, which has so far remained mysterious. Here we
report statistical results of X-ray flares of GRBs with known
redshifts, and show that X-ray flares and solar flares share three
statistical properties: power-law energy frequency distributions,
power-law duration-time frequency distributions, and power-law
waiting time distributions. All of the distributions can be well
understood within the physical framework of a magnetic
reconnection-driven self-organized criticality system. These
statistical similarities, together with the fact that solar flares
are triggered by a magnetic reconnection process taking place in the
atmosphere of the Sun, suggest that X-ray flares originate from
magnetic reconnection-driven events, possibly involved in
ultra-strongly magnetized millisecond pulsars\cite{f10,f11}.
\end{abstract}

Gamma-ray bursts (GRBs) are flashes of gamma-rays occurring at the
cosmological distances with an isotropic-equivalent energy release
from $10^{51}$ to $10^{54}$ ergs\cite{f9,z07,f12,f13}. They can be
sorted into two classes: short-duration hard-spectrum bursts ($<2$
s) and long-duration soft-spectrum bursts\cite{f14}. Thanks to the
rapid-response capability and high sensitivity of the \emph{Swift}
satellite\cite{f15}, numerous unforeseen features have been
discovered, one of which is that about half of bursts have large,
late-time X-ray flares with short rise and decay times\cite{f4,f5}.
The unexpected X-ray flares with an isotropic-equivalent energy from
$10^{48}$ to $10^{52}$ ergs have been detected for both long and
short bursts\cite{f4,f6,f7}. The occurrence times of X-ray flares
range from a few seconds to $10^6$ seconds after the GRB
trigger\cite{f8}. Until now, the physical origin of X-ray flares has
remained mysterious, although some models have been
proposed\cite{f9,z07}. Due to 8-year observations of \emph{Swift},
plentiful of X-ray flare data has been collected. Here we
investigate the energy release frequency distribution, duration-time
frequency distribution and waiting time distribution of GRB X-ray
flares for the first time. On the other hand, it is well known that
solar flares with timescale of hours are explosive phenomena in the
solar atmosphere with energy release of about $10^{28}-10^{32}$
ergs, which are widely believed to be triggered by a magnetic
reconnection process\cite{f16}. They have been observed in broadband
electromagnetic waves, but we here focus on solar hard X-ray flares.

Although X-ray flares are common phenomena in GRBs and the Sun, the
flare energy spans about 20 orders of magnitude and an outstanding
question appears, i.e., do GRB X-ray flares and solar flares have a
similar physical mechanism? Interestingly, some theoretical models
have suggested that GRB X-ray flares could be powered by magnetic
reconnection events\cite{f10,f11}. However, a physical analogy
between GRB X-ray flares and solar flares has not yet been
established.

We search for statistical similarities between GRB X-ray flares and
solar flares. In particular, we compare statistical properties of
the energy release frequency, duration time and waiting time
distributions of GRB X-ray flares and solar flares. For X-ray flares
of GRBs with known redshifts, we employ the published and archival
observed data that allow us to estimate the energy release, duration
time and waiting time of each X-ray flare\cite{f4,f5,f6,f7,f8}. The
total number of flares is 83, including 9 short-burst flares and 74
long-burst flares. The isotropic energy of one flare in the 0.3-10
keV band can be calculated by $E_{\rm iso}=4\pi d^2_L(z)S_F/(1+z)$,
where $S_F$ is the fluence, and $d_L(z)$ is the luminosity distance
calculated for a flat $\Lambda$CDM universe with $\Omega_M=0.3$,
$\Omega_\Lambda=0.7$ and $H_0=70$ km\,s$^{-1}$\,Mpc$^{-3}$. The
Malmquist bias depending on the luminosity function is poorly known
at present. Although GRBs spread over a wide range of redshifts, the
Malmquist bias is claimed to be small and negligible\cite{f17}. The
waiting time in the source's rest frame can be obtained by $\Delta
t=(t_{i+1}-t_i)/(1+z)$, where $t_{i+1}$ is the observed starting
time of the $i+1$th flare, $t_i$ is the observed starting time of
the $i$th flare, and $1+z$ is the factor to transfer the time into
the source-frame one. For the first flare appearing in an afterglow,
the waiting time is taken to be $t_1/(1+z)$. We list the measured
parameters of 83 X-ray flares in Table S1.

Figure 1 shows the cumulative energy distribution of GRB X-ray
flares and the energy frequency distribution of solar hard X-ray
flares. If the number of events $N(E)dE$ with energy between $E$ and
$E+dE$ obeys a power-law relation, $N(E)dE\propto E^{-\alpha_E}dE$
for $E<E_{\rm max}$, with index of $\alpha_E$ and cutoff energy of
$E_{\rm max}$, then we calculate the number of events with energy
larger than $E$ through $N(>E)=a+b[E^{1-\alpha_E}-E_{\rm
max}^{1-\alpha_E}]$, where $a$ and $b$ are two parameters. In order
to obtain the best-fitting parameters, the Markov chain Monte Carlo
technique is used in our calculations. We obtain $\alpha_E \simeq
1.06\pm 0.15$ for GRB X-ray flares. In addition, the blue and green
curves in Figure 1 represent the energy frequency distribution with
$\alpha_E=1.65\pm 0.02$ and $1.53\pm0.02$ for RHESSI\cite{f18} and
HXRBS\cite{f19} solar flares, respectively. In Figure 2, we present
the duration-time ($T$) frequency distributions of solar flares and
GRB X-ray flares, which can also be fitted by a power-law relation
with index of $\alpha_T$, i.e. $N(T)dT\propto T^{-\alpha_T}dT$. The
red lines in Figure 2 show a power-law fit with $\alpha_T=-1.10\pm
0.15$ and $-2.00\pm 0.05$ for GRB X-ray flares and solar flares,
respectively.

Although the energy and duration-time frequency distributions for
two kinds of flares are apparently different, we next show that
these distributions can be well understood within one physical
framework. The energy and duration frequency distributions of solar
fares have been thought to be attributed to a magnetic reconnection
process based on the fractal-diffusive avalanche
model\cite{f18,f20,f22}. We further discuss this model to explain
the energy and duration-time frequency distributions of GRB X-ray
flares. For a self-organized criticality (SOC) avalanche, due to
diffuse random walking, a statistical relationship\cite{f20} between
size scale $L$ and duration time $T$ of the avalanche is $L\propto
T^{1/2}$, and a probability distribution of size $L$ is argued as
$N(L)dL\propto L^{-S}dL$ for the three Euclidean dimensions $S=1$, 2
and 3. This probability argument is based on the assumption that the
occurrence frequency or number of events is equally likely
throughout the system. So the index of the duration frequency
distribution of flares is given by\cite{f20}
\begin{equation}
\alpha_T=\frac{S+1}{2}.
\end{equation}
This index becomes $\alpha_T=1$ for $S=1$ and $\alpha_T=2$ for
$S=3$, which can well explain the observed duration distributions of
GRB X-ray flares and solar flares. On the other hand, the index of
the energy frequency distribution can be written by\cite{f20}
\begin{equation}
\alpha_E=\frac{3(S+1)}{S+5}.
\end{equation}
It is easy to see that the index $\alpha_E=1$ for $S=1$ and
$\alpha_E=1.5$ for $S=3$, which are remarkably consistent with the
observed indices of GRB X-ray flares and solar flares. A power-law
distribution of occurrence frequency is characteristic of the SOC
system\cite{f18}. According to equations (1) and (2), the power-law
indices of the energy and duration-time distributions of SOC depend
on the fractal geometry of the energy dissipation domain\cite{f20}.
Thus, it is clear to find that GRB X-ray flares and solar flares
correspond to the one-dimension ($S=1$) and three-dimension ($S=3$)
cases, respectively. Therefore, GRB X-ray flares and solar flares
share energy and duration-time frequency distributions, suggesting
that they have a similar physical origin.

Now we discuss the waiting time distributions for two kinds of
flares. The waiting time is defined as the time interval between two
successive events. Its distribution tells us information on whether
events occur as independent events, and provides the mean rate of
event occurrence\cite{f18}. It has been suggested that the waiting
time distribution of solar flares can be described by a power-law
distribution with index of about $-2.0$ for long waiting
times\cite{f21,f23}. But the waiting time distribution of GRB X-ray
flares has not been studied before. Figure 3 displays the waiting
time distribution of GRB X-ray flares and solar flares. Excluding
the fluctuations of short waiting times, the waiting time
distribution of GRB X-ray flares is also a power-law with index
$-1.80\pm0.20$. The solar flares with waiting time larger than about
2 hours observed by RHEESI during 2002-2009 can be fitted by a
power-law function with an index of $-2.0\pm 0.05$. Thus, GRB X-ray
flares and solar flares have similar waiting time distributions,
which can be explained by non-stationary Poisson
processes\cite{f21}. A Poissonian random process has an exponential
waiting time distribution for a stationary flare rate and a
power-law-like waiting time distribution for a non-stationary flare
rate, which is the predication of the SOC theory\cite{f18}. For a
non-stationary Poisson process, the waiting time distribution can be
expressed by\cite{f21}
\begin{equation}
P(\Delta t) = {\lambda_0 \over (1 + \lambda_0 \Delta t)^2},
\end{equation}
where $\lambda_0$ is the mean rate of flares. For large waiting
times $(\Delta t \gg 1/\lambda_0)$, equation (3) approaches a
power-law relation $P(\Delta t)\approx (1/\lambda_0)(\Delta
t)^{-2}$, which is consistent with the observations. We can see from
Figure 3 that the breakpoint is around $\Delta t_0=1/\lambda_0\sim
20$ s in the source's rest frame for GRB X-ray flares (so the mean
rate is $\lambda_0\sim 0.05\,{\rm s}^{-1}$), and $\Delta t_0\sim
1.2$ hrs for solar flares.

The statistical similarities between GRB X-ray flares and solar
flares suggest a similar physical origin, i.e., magnetic
reconnection. When and where does the magnetic reconnection event
take place for an X-ray flare? The observations imply that the
central engines of GRBs have long-lasting
activity\cite{f2,f4,f5,f6,f7,f8} and X-ray flares arise from late
internal shocks\cite{f24,f25}, which could be formed through
collisions of shells ejected after the prompt gamma-ray emission has
ended. This implication is based on two following facts: first, the
short rise and decay timescales and corresponding distributions of
X-ray flares require that the central engines restart at late
times\cite{f26}, and second, the peak time of an X-ray flare
observed by {\em Swift} is nearly equal to the ejection time of a
relativistic outflow from the central engine if the decaying phase
of the flare is understood as being due to the high latitude
emission from the outflow\cite{f27}. Therefore, the magnetic
reconnection event of an X-ray flare should occur at late times.
Such an event could be powered by a differentially rotating,
ultra-strongly magnetized, millisecond pulsar after the merger of a
neutron star-neutron star binary or the collapse of a massive
star\cite{f10}. The differential rotation leads to windup of
poloidal magnetic fields in the interior and the resulting toroidal
fields are strong enough to float up and break through the stellar
surface\cite{f28,f29}. Magnetic reconnection-driven multiple
explosions then occur, producing X-ray flares. Because these
explosions usually take place around the pulsar's surface along the
rotation axis\cite{f28,f29}, a cylinder-like magnetic-reconnection
region with height $L$ has a volume $\sim \pi R_{\rm P}^2L$ (where
$R_{\rm P}$ is the pulsar's radius) and thus the probability
distribution $N(L)$ is inversely proportional to $L$. This implies
that GRB X-ray flares belong to the one-dimension ($S=1$) SOC case,
as we found above. A variant of the magnetic reconnection mechanism
is internal dissipation of relativistic winds from postburst
millisecond magnetars (a type of pulsar with magnetic field strength
of $\sim 10^{14}-10^{15}\,$Gauss), which could account for the
observational properties of GRB X-ray flares\cite{f11}.

We have suggested the magnetic reconnection mechanism as the
physical origin of GRB X-ray flares based on three statistical
similarities between X-ray flares and solar flares. We found that
such magnetic reconnection-driven events correspond to the $S=1$ SOC
case for GRB X-ray flares. This is different from solar flares,
which are thought to be due to a magnetic reconnection-driven $S=3$
SOC process\cite{f20}. Our work has at least three implications.
First, it could not only help to understand the central engines of
GRBs, but also help to study applications of solar
magnetic-reconnection theories. Second, it could stimulate numerical
simulations on a magnetic reconnection-driven self-organized
criticality process under extreme astrophysical situations (e.g.,
ultra-strongly magnetized millisecond pulsars). Third, it could
bring about similar statistical studies of other astrophysical
explosive phenomena.

\noindent {\bf Acknowledgements} We thank M. J. Aschwanden, P. F.
Chen, Y. Dai, M. D. Ding, Y. Guo, Y. F. Huang, and X. Y. Wang for
discussions. This work was supported by the National Natural Science
Foundation of China (grant No. 11103007 and 11033002).

\noindent {\bf Author Contributions} F.Y.W. analyzed the
observational data and explained the statistical results based on a
self-organized criticality theory. Z.G.D. suggested such an analysis
and proposed the physical origin of X-ray flares. Both authors wrote
this paper together.

\noindent {\bf Author Information} The authors declare that they
have no competing financial interests. Correspondence and requests
for materials should be addressed to Z.G.D. (dzg@nju.edu.cn) or
F.Y.W. (fayinwang@nju.edu.cn).

\clearpage
\begin{figure}
\includegraphics[width=\textwidth]{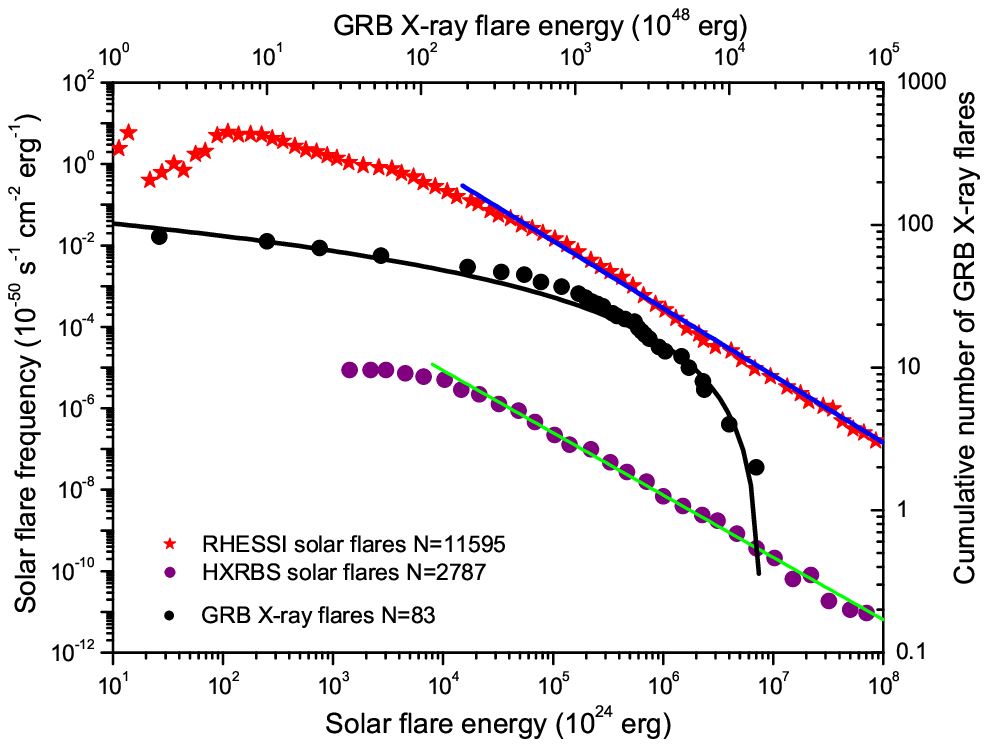}
\caption{\label{Fig1} The cumulative energy distribution of GRB
X-ray flares and the energy frequency distribution of solar hard
X-ray flares. We employ 11595 solar flares\cite{f18} shown as red
stars from RHESSI during 2002-2007, and 2787 flares\cite{f19} shown
as purple dots from HXRBS during 1980-1982. The black curve gives
the cumulative energy distribution $N(>E)=a+b[E^{1-\alpha_E}-E_{\rm
max}^{1-\alpha_E}]$. Using the Markov chain Monte Carlo fitting
technique, we obtain $\alpha_E\sim 1.06\pm 0.15$ for GRB X-ray
flares. For a differential frequency distribution $N(E)dE\propto
E^{-\alpha_E}dE$ of solar flares, the blue and green curves give
$\alpha_E=1.65\pm 0.02$ and $1.53\pm0.02$ for the RHESSI and HXRBS
samples, respectively. The magnetic reconnection process predicts
that the value of $\alpha_E$ is $1.0$ and $1.5$ for one fractal
dimension and three fractal dimensions, respectively\cite{f20}.
Although the power-law indices are apparently different for GRB
X-ray flares and solar flares, their physical mechanisms are
similar, as discussed in the text.}
\end{figure}

\clearpage
\begin{figure}
\includegraphics[width=\textwidth]{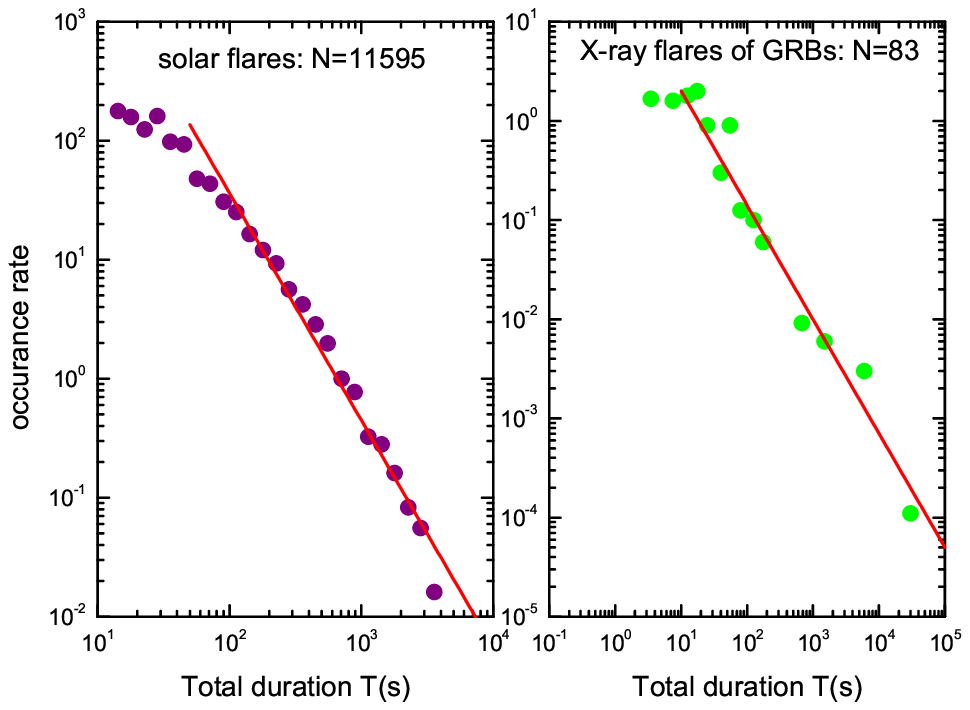}
\caption{\label{Fig2} The relation between the occurrence rate and
duration time for solar flares (left) and GRB X-ray flares (right).
The occurrence rate is defined by the ratio of the number of X-ray
flares in the bin to the bin width. The total duration times of
11595 solar flares\cite{f18} were observed by RHESSI during
2002-2007. The best-fit power-law indices are $-2.00\pm 0.05$ and
$-1.10\pm 0.15$ for solar flares and GRB X-ray flares, respectively.
}
\end{figure}

\clearpage
\begin{figure}
\includegraphics[width=\textwidth]{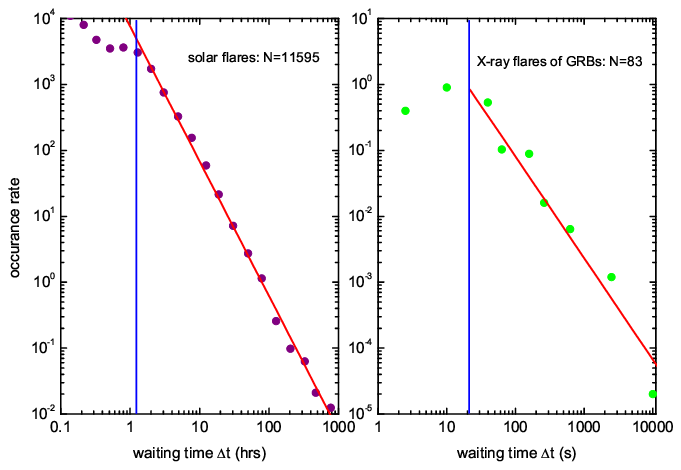}
\caption{\label{Fig2} The relation between the occurrence rate and
waiting time for solar flares (left) and GRB X-ray flares (right).
The waiting times of 11595 solar flares\cite{f21} were observed by
RHESSI during 2002-2007. The best-fit power-law indices are
$-1.80\pm 0.20$ and $-2.04\pm 0.03$ for GRB X-ray flares and solar
flares with large waiting times, respectively. The blue vertical
lines show the waiting time at the breakpoint, corresponding to the
mean rate of flares. The waiting times at the breakpoint are about
20 s in the source's rest frame and 1.2 hrs for GRB X-ray flares and
solar flares, respectively.}
\end{figure}

\clearpage

\begin{longtable}{ccccccccccc}
\multicolumn{11}{l}{\large{{\bf Supplementary Information}}} \\
\multicolumn{11}{l}{{\bf Table~S1.} The measured parameters of X-ray flares of gamma-ray bursts.} \\
\hline\hline Name & $z$ & $t_{\rm start}^{\rm a}$ & $T$$^{\rm b}$ & $S_F$ & $E_{\rm iso}^{\rm c}$ & Ref. \\
 GRB &  & (s) & (s) & ($10^{-8}$\,erg\,cm$^{-2}$) & ($10^{50}$\,ergs) &  &  \\
 \hline
050730 & 3.967 & 190.5 & 42.9 & 2.9 & 8.82 & \cite{f6} \\
050730 & 3.967 & 311.6 & 110.7 & 10 & 30.4 & \cite{f6} \\
050730 & 3.967 & 606.9 &98.4 & 3.7 & 11.25 & \cite{f6} \\
050908 & 3.344 & 362.4 & 115.9 & 1.9 & 4.4  & \cite{f6} \\
060115 & 3.53 & 369.7 & 83.4& 0.6 & 1.52  & \cite{f6} \\
060210 & 3.91 & 136.5 & 60.7 & 23 & 68.40 & \cite{f6} \\
060210 & 3.91 & 350.9 & 60.2 & 12 & 35.67 & \cite{f6} \\
060418 & 1.489 & 122.3 & 25.1 & 48 & 27.01 & \cite{f6} \\
060512 & 0.4428 & 167.4 & 79.3 & 4.5 & 0.22 & \cite{f6} \\
060526 & 3.221 & 96.2 & 24.5 & 32 & 69.65 & \cite{f6} \\
060526 & 3.221 & 257.5 & 37.7 & 25 & 54.41 & \cite{f6} \\
060526 & 3.221 & 279.5 & 49.8 & 27 & 58.77 & \cite{f6} \\
060526 & 3.221 & 316.8 & 71.9 & 13 & 28.30 & \cite{f6} \\
060604 & 2.68 & 116.1 & 32.0 & 13 & 20.84 & \cite{f6} \\
060604 & 2.68 & 163.7 & 21.9 & 7.9 & 12.67 & \cite{f6} \\
060607A& 3.082 & 92.1 & 15.9 & 4.4 & 8.91 & \cite{f6} \\
060607A& 3.082 & 193.0 & 75.5 & 24 & 48.59 & \cite{f6} \\
060707 & 3.425 & 174.5 & 34.6 & 0.7 & 1.68 & \cite{f6} \\
060714 & 2.711 & 55.6 & 8.2 & 3.8 & 6.21 & \cite{f6} \\
060714 & 2.711 & 103.9 & 49.6 & 33 & 53.94 & \cite{f6} \\
060714 & 2.711 & 131.5 & 14.0 & 9.5 & 15.53 & \cite{f6} \\
060714 & 2.711 & 151.5 & 21.1 & 11 & 17.98 & \cite{f6} \\
060729 & 0.54 & 163.4 & 43.7 & 93 & 6.99 & \cite{f6} \\
060814 & 0.84 & 119.5 & 39.0 & 31 & 5.73 & \cite{f6} \\
060904B& 0.703 & 125.9 & 78.5 & 200 & 25.81 & \cite{f6} \\
060908 & 1.8836 & 494.8 & 151.2 & 0.4 & 0.35 & \cite{f6} \\
060908 & 1.8836 & 605.6 & 178.9 & 0.5 & 0.43 & \cite{f6} \\
070306 & 1.4959 & 169.7 & 43.4 & 21 & 11.92 & \cite{f6} \\
070318 & 0.836 & 179.0 & 28.9 & 1.0 & 0.18 & \cite{f6} \\
070318 & 0.836 & 203.7 & 147.9 & 13.0 & 2.38 & \cite{f6} \\
070721B & 3.626 & 256.2 & 121.7 & 6.0 & 15.82 & \cite{f6} \\
070721B & 3.626 & 297.8 & 11.7 & 2.6 & 6.85 & \cite{f6} \\
070721B & 3.626 & 328.7 & 17.2 & 1.9 & 5.01 & \cite{f6} \\
070721B & 3.626 & 575.0 & 242.0 & 2.2 & 5.80 & \cite{f6} \\
070724A & 0.457 & 30.3 & 32.3 & 2.8 & 0.15 & \cite{f6} \\
070724A & 0.457 & 136.9 & 85.7 & 2.0 & 0.11 & \cite{f6} \\
071031 & 2.692 & 66.6 & 61.2 & 9.5 & 15.35 & \cite{f6} \\
071031 & 2.692 & 181.1 & 44.0 & 8.0 & 12.92 & \cite{f6} \\
071031 & 2.692 & 242.8 & 51.9 & 5.4 & 8.72 & \cite{f6} \\
071031 & 2.692 & 352.7 & 276.1 & 19.0 & 30.69 & \cite{f6} \\
080210 & 2.641 & 153.9 & 35.7 & 9.3 & 14.54 & \cite{f6} \\
080310 & 2.42 & 463.5 & 161.7 & 29.0 & 39.06 & \cite{f6} \\
080310 & 2.42 & 526.5 & 59.6 & 18.0 & 24.25 & \cite{f6} \\
050416A & 0.654 & 1.5E6 & 5.0E5 & 3.4 & 0.38 & \cite{f8} \\
060223A & 4.41 & 891 & 293.0 & 0.9 & 3.23 & \cite{f8} \\
060223A & 4.41 & 1360 & 142.2 & 0.5 & 1.79 & \cite{f8} \\
060906 & 3.69 & 908 & 5804 & 1.2 & 3.25 & \cite{f8} \\
060906 & 3.69 & 6200 & 1459.0 & 0.5 & 1.36 & \cite{f8} \\
070318 & 0.836 & 1.2E5 & 1.12E5 & 2.9 & 0.53 & \cite{f8} \\
071031 & 2.692 & 4120 & 1745 & 0.2 & 0.32 & \cite{f8} \\
071112C & 0.82 & 296 & 187.5 & 1.8 & 0.32 & \cite{f8} \\
071112C & 0.82 & 810 & 224.9 & 1.2 & 0.21 & \cite{f8} \\
090417B & 0.35 & 1250 & 275.1 & 47.6 & 1.46 & \cite{f8} \\
090417B & 0.35 & 1420 & 217.8 & 45.1 & 1.38 & \cite{f8} \\
090417B & 0.35 & 1620 & 390.3 & 47.3 & 1.45 & \cite{f8} \\
090809 & 2.74 & 2480 & 2667 & 6.9 & 11.48 & \cite{f8} \\
050724 & 0.258 & 230 & 65.11 & 4.41 & 0.072 & \cite{f7} \\
050724 & 0.258 & 9630 & 6.46E4 & 8.5 & 0.14 & \cite{f8} \\
070724A & 0.457 & 75 & 22.28 & 0.95 & 0.0507 & \cite{f7} \\
070724A & 0.457 & 90 & 19.1 & 1.26 & 0.067 & \cite{f7} \\
070724A & 0.457 & 150 & 87.8 & 1.83 & 0.097 & \cite{f7} \\
071227 & 0.383 & 150 & 27.55 & 0.55 & 0.0203 & \cite{f7} \\
100117A & 0.92 & 130 & 42.77 & 0.98 & 0.2143 & \cite{f7} \\
100117A & 0.92 & 164 & 67.78 & 2.14& 0.4745 & \cite{f7} \\
100117A & 0.92 & 200 & 22.11 & 0.40 & 0.0878 & \cite{f7} \\
050802 & 1.434 & 312 & 145 & 0.2 & 0.10 & \cite{f4} \\
050814 & 5.3 & 1133 & 841 & 0.2 & 0.95 & \cite{f4} \\
050814 & 5.3 & 1633 & 944 & 0.4 & 1.90 & \cite{f4} \\
050819 & 2.5 & 56 & 197 & 1.9 & 2.71 & \cite{f4} \\
050819 & 2.5 & 9094 & 27628 & 1.0 & 1.42 & \cite{f4} \\
050820A & 2.617 & 200 & 182 & 68.1 & 104.8 & \cite{f4} \\
050904 & 6.29 & 343 & 227 & 23.8 & 145.5 & \cite{f4} \\
050904 & 6.29 & 857 & 284 & 1.6 & 9.78 & \cite{f4} \\
050904 & 6.29 & 1149 & 194 & 1.0 & 6.11 & \cite{f4} \\
050904 & 6.29 & 5085 & 3916 & 8.5 & 51.96 & \cite{f4} \\
050904 & 6.29 & 16153 & 8713 & 10.7 & 65.41 & \cite{f4} \\
050904 & 6.29 & 18383 & 20230 & 5.7 & 34.85 & \cite{f4} \\
050904 & 6.29 & 25618 & 5360 & 4.1 & 25.06 & \cite{f4} \\
050915A & 2.53 & 55 & 115 & 2.7 & 3.93 & \cite{f4} \\
060108 & 2.03 & 193 & 236 & 0.2 & 0.20 & \cite{f4} \\
060108 & 2.03 & 4951 & 33035 & 7.0 & 6.93 & \cite{f4} \\
060124 & 2.3 & 283 & 361 & 271.3 & 334.6 & \cite{f4} \\
060124 & 2.3 & 644 & 363 & 124.0 & 153.0 & \cite{f4} \\
\hline
\end{longtable}
Note: (a) In this table, the waiting time in the source frame can be
obtained by $\Delta t=(t_{i+1}-t_i)/(1+z)$, where $t_{i+1}$ is the
observed starting time of the $i+1$th flare, and $t_i$ is the
observed starting time of the $i$th flare. For the first flare in
one GRB, the waiting time is $t_1/(1+z)$. (b) The flare duration
time in the observer frame is $T$, and the duration time in the
source frame is $T/(1+z)$. These flares are clearly distinguishable
from the underlying continuum emission\cite{f4,f5,f6,f7,f8}, so the
duration suffers no bias. (c) The flare isotropic energy is
calculated by $E_{\rm iso}=4\pi d^2_L(z)S_F/(1+z)$, where $S_F$ is
the fluence, and $d_L(z)$ is the luminosity distance calculated for
a flat $\Lambda$CDM universe with $\Omega_M=0.3$,
$\Omega_\Lambda=0.7$ and $H_0=70$ km s$^{-1}$ Mpc$^{-3}$.

\end{document}